\documentclass[runningheads]{llncs}
\usepackage{graphicx}
\usepackage{tabularx}
\usepackage[utf8]{inputenc}

\begin{document}
\title{Validating psychometric survey responses\thanks{To university Columbia, the project Capstone and dotin Inc. for provided data sources.}}
%
%
\author{Alberto Mastrotto\inst{1}\and
Anderson Nelson\inst{1}\and
Dev Sharma\inst{1}\and
Ergeta Muca\inst{1} \and \\
Kristina Liapchin\inst{1} \and
Luis Losada\inst{1}\and
Mayur Bansal\inst{1}\and\\
Roman S. Samarev\inst{2, 3}\orcidID{0000-0001-9391-2444}
}
\authorrunning{A.Mastrotto et al.}
%
\institute{Columbia University, 116th St and Broadway, New York, NY 10027, USA\\
	\email{an2908@columbia.edu} \\
	\url{https://www.columbia.edu/} \and
	dotin Inc, Francisco Ln. 194, 94539, Fremont CA, USA\\
	\email{romans@dotin.us}
	\url{https://dotin.us} \and
	Bauman Moscow State Technical University, ul. Baumanskaya 2-ya, 5/1, 105005, Moscow, Russia, \url{https://bmstu.ru/en} 
}

\maketitle              
\begin{abstract}
We present an approach to classify user validity in survey responses by using a machine learning techniques. The approach is based on collecting user mouse activity on web-surveys and fast predicting validity of the survey in general without analysis of specific answers. Rule based approach, LSTM and HMM models are considered. The approach might be used in web-survey applications to detect suspicious users behaviour and request from them proper answering instead of false data recording.

\end{abstract}

\section{Introduction}

Survey responses can be a crucial data point for researchers and organization seeking to gain feedback and insight. Modern survey design incentives users to complete as many surveys as possible in order to be compensated, in some situations, users are falsifying the response, thus rendering the response invalid. Organization and researchers can reach the wrong conclusion if the user responses are largely invalid. Mouse and keyboard are most common controls available for PC users. Even now, with plenty of touch screen devices, from programmatic point of view, touch screen generates mouse related commands. We gathered mouse data tracking and created features on: Time, Screen coverage, Distance traveled, and Direction of movements.  The basis of creating these features was on the literature review of mouse path analytic as well as common business knowledge. Although not all features ended up being used in our final models, they played a big role in our exploratory data analysis and in developing our models to help us get the best and most accurate results. A detailed table containing all features created and used in modeling can be found in Figure 1 in the Appendix.

\section{Related works}
Using of machine learning approaches with all available for collection data is very common approach for researchers last years. We found different directions of research of mouse tracks: mood analysis, authentication based on user specific analysis, common behaviour analysis.

One of early works related to emotion analysis \cite{DBLP:journals/eaai/KaklauskasZSDSSSPMKBIG11} considered a special prepared mouse with additional sensors like electrogalvanic skin conductance, temperature, humidity and pressure sensors. But their mouse events subsystem calculated speed of mouse pointer’s movement, acceleration of mouse pointer’s movement, amplitude of hand tremble, scroll wheel use right- and left-click frequency, idle time. The authors use these values in their common regression model, but there are no correlations presented in term of exact mouse movement use.

The work \cite{Motwani-15} demonstrates use of multimodal user identification based on keyboard and mouse activity. The authors used False Rejection Rate as a quality value and show it $\approx3.2\%$. Main features their used for mouse analysis: traveled distance between clicks, time intervals between releasing and next pressing, and vice versa, double click values like times, time interval, distance, and similar drag-and-drop parameters.

A little bit simplified approach for a user authentication was shown in the paper \cite{Singh-11}. Here, only distance travelled by the mouse was used. And two hypotheses were considered: mouse speed increases with the distance travelled, mouse speed is different in different directions considerably. The key idea was to restrict the screen for mouse activity recording by a set of 9 buttons placed inside a square. The control parameters were used false acceptance and false rejection rates (FAR and FRR) with $1.53$ and $5.65$ maximum values respectively. 

The paper \cite{Suganya2016ImprovingTP}, also, describes an approach for user authentication, but mouse extracted features are operation frequency, silence ratio as a percent of idle time, movement time and offsets, average movement time and distance, distribution of cursor positions, horizontal, vertical, tangential velocity, acceleration and jerk, slope angle and curvature. Dimensional reduction was implemented with diffusion map algorithm. And relationship between heat diffusion and random walk Markov chain was calculated. Diffusion distances were used in a Hopfield network based classifier. The results were shown as $FAR\approx5.05$ and $FRR\approx4.15$.

Later work \cite{10.1145/3055635.3056620} uses multiple classifiers for solving the same task of user authentication and demonstrates better results $FAR\approx0.064$ and $FRR\approx0.576$. Their features in mouse tracking analysis were total number of point for a certain interval, total amount of time when mouse movement was in delay, how many times the Trajectory was in delay, number of action, total Length and STDEV of the Trajectory Length and Slope, curvature as number of changes between the angles and total length of the Trajectory. The authors used SVM, K Nearest Neighbor and Naïve Bayes classifiers.

The paper \cite{Salmeron-Majadas-ML-MK} is devoted to user specific behaviour on keyboard and mouse use. The authors used following set of mouse related features: distance, speed, acceleration, direction and angle, element clicks, click duration and scroll, and pauses. For data collection the tool MOKEETO was developed, and that tool provided both mouse and keyboard related events. The authors used SMOTE oversampling and PCA for preprocessing. And decision trees, random forest, support vector machine, and Naïve Bayes classifiers. The results demonstrate ability to differentiate users behaviour but there are no separate mouse and keyboard features investigation were shown.

The paper \cite{Elbahi-16} considered use of mouse movement for e-learning activities recognition. In the paper Possibilistic Hidden Markov Model and Possibilistic Conditional Random Fields model approaches were described. The key idea of the paper is to catch an area of interests as a mouse cursor fixation over some image on a screen with the OGAMA tool. The tool gives some tasks and records mouse activity. As features for analysis in that case, the authors used total time of a task, time between two cursor fixations, distance. The authors demonstrate up to 90\% accuracy of a task recognition.

In the paper \cite{Yamauchi-17} a mouse cursor motions analysis for emotion reading was considered. The authors demonstrated a set of images, asked to show which ones are appropriate answers, and recorded a cursor movements. As the authors tried to work in emotional area, they also combined tests with different music, movies and art background. Key features for them in a cursor analysis were attraction and direction changes (zigzag). The authors used SVM method of mouse tracks analysis and were able to recognize only some of common emotions.

\section{Methods selection}
Based on our initial exploratory data analysis, we proceeded with building a few different models to help us identify fraudulent survey responses with the goal of improving the current validation method used by dotin Inc. We developed the following three methods throughout the course of this project:
\begin{itemize}
\item{Rule-based Approach}
\item{Long Short-Term Memory (Supervised Learning)}
\item{Hidden Markov Model (Unsupervised Learning)}
\end{itemize}
We decided to use these three approaches to compare how the different methods would perform, considering the lack of accurately labeled data in the original dataset. That way, we would be able to make better and more well-informed recommendations for dotin Inc. with regards to a new validation method to use for their psychometric survey responses. 

\section{RULE-BASED APPROACH}
From our EDA we identified that the tracking method used to generate the mouse path dataset presented some challenges as many of the user’s paths weren’t fully recorded. Out of the 755 user’s data, only 54 fulfilled the basic requirement of clicking the 196 radio buttons pertaining to individual questions. Therefore, we added our data collection recommendations in the final section of our paper. 
Before diving into the modeling, our team found essential to create alternative ways to flag anomalous users other than dotin’s current validation method. In order to generate such features, we used both common business sense and advanced outlier detection techniques that allow us to understand each user from different angles. Such features will serve as a way to validate dotin’s current validation method as well as allow us to generate basic business rules to flag suspicious behavior. Some of these features will then be used to test our models.

\subsection{Anomalies by scores}
From our analysis, we discovered that 150 of the 755 users surveyed answer at least one page of the survey with all of the same scores. We then assume that there is no page where such an event would be plausible, therefore these users are flagged as suspicious.

\begin{figure}[h]
  \centering
  \includegraphics[width=\linewidth]{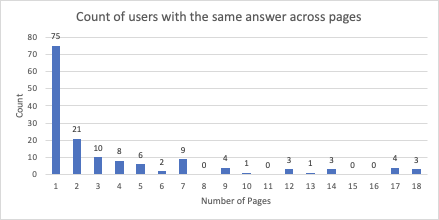}
  \caption{Count of users with the same answer across pages}
\end{figure}

\subsection{Anomalies by time}
We then proceeded to focus on the time perspective by estimating the read time that an honest user would take to read the survey and compared it with the actual completion time taken by each individual user. The benchmark read time of a regular user was derived from Medium’s read-time algorithm, which is based on the average reading speed of an adult (~256 wpm). The read time was calculated for all the individual questions in our users’ surveys and compared to the time it took them to click one radio button to another (an indication of them moving from one question to the other). From our analysis, on average, a user that completed the entire survey would need 5 minutes and 30 seconds to at least read all the 196 questions, yet  33\% of our surveyed users took less time than that. Therefore, we flagged users that take less than the calculated reading times as anomalous.

\subsection{Anomalies by topic}
Finally, we focused on the first 40 questions of the survey to create our own topics and scored each user based on how they deviate in answering the survey questions. For each topic, we aggregated questions that are either positive or negative (i.e. Tidy/Untidy) and we analyzed how users answer differently for similar questions. The underlying assumption is that if the user is deviating from their answers each time, this indicates that he/she is not fully paying attention to the questions. Questions with opposite behavioral traits should then present scores that are opposite (low standard deviation). In our analysis, we chose a threshold for a standard deviation of 2 to identify unfocused users, consequently resulting in 33\% of users answering opposite questions with similar answers, (i.e. $Tidy = 5$; $Untidy = 5$). Based on this analysis, such users will then be flagged as suspicious. 

\subsection{Aggregated Flag Scores based on Rule-Based Approach}
In order to identify our suspicious users based on these 3 features, we assign a flag score to each user. This flag score indicates the level of suspicion that our rule-based approach suggests. The value of the flag score ranges from 0 to 1 where 0 indicates that the user can be validated and a value greater than 0 means that the user appears as an anomaly in at least one feature, which suggests that the user is a red flag. Based on our results, we decided to select as outliers all users with a flag score > 0, consequently identifying 310 users i.e. (44\% of the total users).

\textbf{Generating a new validation variable with Autoencoders:}
Due to the outlier detection nature of the features explained above, we decided to take an unsupervised learning approach to create a new validation method. We used an outlier detection algorithm to create our own labels of valid and non-valid users.  The nature of our dataset then required an approach that could deal with many variables but few observations (704 observations that represent features for each user).

\textbf{Training the Autoencoder:}
In order to train our autoencoder, we handpicked 144 users that based on our analysis had completed the entire survey and whose mouse activity data was clean. The autoencoder model trained on these users had 25 Neurons on the input and output layers and two hidden layers of 2 neurons each. The compression used a sigmoid activation function and the mean squared error of the process was 11.49.
The results showed that 76\% of the users were classified as non-outliers while the rest were classified as outliers. Although this method did take into account mouse behavior, we wanted to focus on mouse movement at a more granular level. We further use the output of this validation method as a dependent variable in our LSTM model.

\section{LONG SHORT-TERM MEMORY (LSTM)}
Recurrent Neural Networks (RNN) have grown to be a popular tool in Natural Language Processing for Language Modeling. Hence, RNN implementations are no strangers to sequence-based applications. As in language modeling, an RNN is responsible for predicting the next token. Our approach to applying RNNs to the problem at hand consists of two key stages:
\begin{itemize}
\item{Training a model that can predict a user’s next movement.}
\item{Transferring the learning from the first model to a classifier model for predicting survey response validation trained using autoencoders.}
\end{itemize}

\subsection{Data Preparation}

In order to feed an RNN, we needed to transform our data into a sequential format that the RNN can understand. For this purpose, we created string-based tokens which identified the cardinal directions and magnitudes of a user’s movements. Page changes are identified with the “pagechange” token. All of a user’s movements were appended to a single tokenized list of strings. For example, a user’s movements might start off as [“nw”, “1”, “sw”, “3” …. “pagechange” “ne”, “2”]. For memory efficiency, movements were averaged out between radio clicks. 

Since our RNN’s loss function would be Cross-Entropy instead of Mean Squared Error, we scaled the magnitudes significantly to create large bins. This means that if our model predicts “8” as a magnitude whereas it should have predicted “7” for example, it is justified to penalize the model just as if it would have predicted a “2” because a one-point shift in magnitude is quite significant. Lastly, we split our data into training and validation sets based on a 70:30 split respectively.

\subsection{Model Architecture}

We used Long Short-Term Memory (LSTM) as a model as they are robust against the vanishing gradient problem. Similar to RNNs, our models carried two types of parameters: token embeddings and hidden states. Weights also included those which the LSTM uses to determine how significant of an adjustment should be made for the new sequential input. Tokenized user movements were inputted in mini-batches of 8 and trained on a 6GB 1070 GPU. For batches of a user where input length differed, padding was added to the end of shorter sequences. We used the cross-entropy loss function, and the evaluation metric for both the language model and the classifier was Accuracy. 
Once the first model was trained, we replaced the final linear layer with a classification head of N x 2 dimensions, which produced a binary label where N is the input dimensions of the final hidden state from our LSTM. 

\subsection{Results}
In stage 1 of the language model, we trained the model on our training set. We achieved the following results in predicting the next token on our validation set.

\begin{table}
  \centering
  \caption{LSTM stage 1 results}
  \label{tab:commands}
  \begin{tabular}{|c|c|c|c|c|}
    \hline
epoch  &  train\_loss & valid\_loss & accuracy & time \\
    \hline
0  &  2.336163  &  1.453722 &   0.521071 &   00:09\\
1  &  1.721388  &  1.280523 &   0.577649 &   00:09\\
2  &  1.470130  &  1.229108 &   0.586503 &   00:09\\
3  &  1.350252  &  1.212001 &   0.588408 &   00:09\\
4  &  1.275603  &  1.179907 &   0.595714 &   00:09\\
5  &  1.218212  &  1.168359 &   0.584360 &   00:09\\
… &……… &……… &……… &………\\
20 &   0.956453 &   0.997654 &   0.638661 &   00:09\\
21 &   0.945954 &   0.995324 &   0.638824 &   00:09\\
22 &   0.944055 &   0.996780 &   0.639673 &   00:09\\
23 &   0.939271 &   0.996102 &   0.640015 &   00:09\\
24 &   0.938368 &   0.995273 &   \textbf{0.639896} &   00:09\\
    \hline
  \end{tabular}
\end{table}

We received an accuracy of $\sim64\%$ after twenty-five epochs. Now that we have developed a model that was able to predict the next word, we removed the language model head and replaced it with a classifier head with randomly generated parameters. Hence, we trained this head to classify the validation status of surveys. Following are the results of predicting the survey validation status after five epochs:

\begin{table}
  \centering
  \caption{LSTM stage 2 results}
  \label{tab:commands}
  \begin{tabular}{|c|c|c|c|c|}
    \hline
epoch  &  train\_loss & valid\_loss & accuracy & time \\
    \hline
0 &   0.716256 &   0.499177 &   0.881944&    00:16\\
1 &   0.649158 &   0.408814 &   0.840278&    00:16\\
2 &   0.583171 &   0.345984 &   0.902778&    00:16\\
3 &   0.524735 &   0.313288 &   0.895833&    00:16\\
4 &   0.498957 &   0.332085 &   \textbf{0.895833}&    00:16\\
    \hline
  \end{tabular}
\end{table}

The LSTM  produced a $\sim90\%$ accuracy on predicting whether a user’s survey response is valid or invalid. Following is the confusion matrix and the classification report:

\begin{figure}[h]
  \centering
  \includegraphics[
  width=15cm,
  height=6cm,
  keepaspectratio,
]{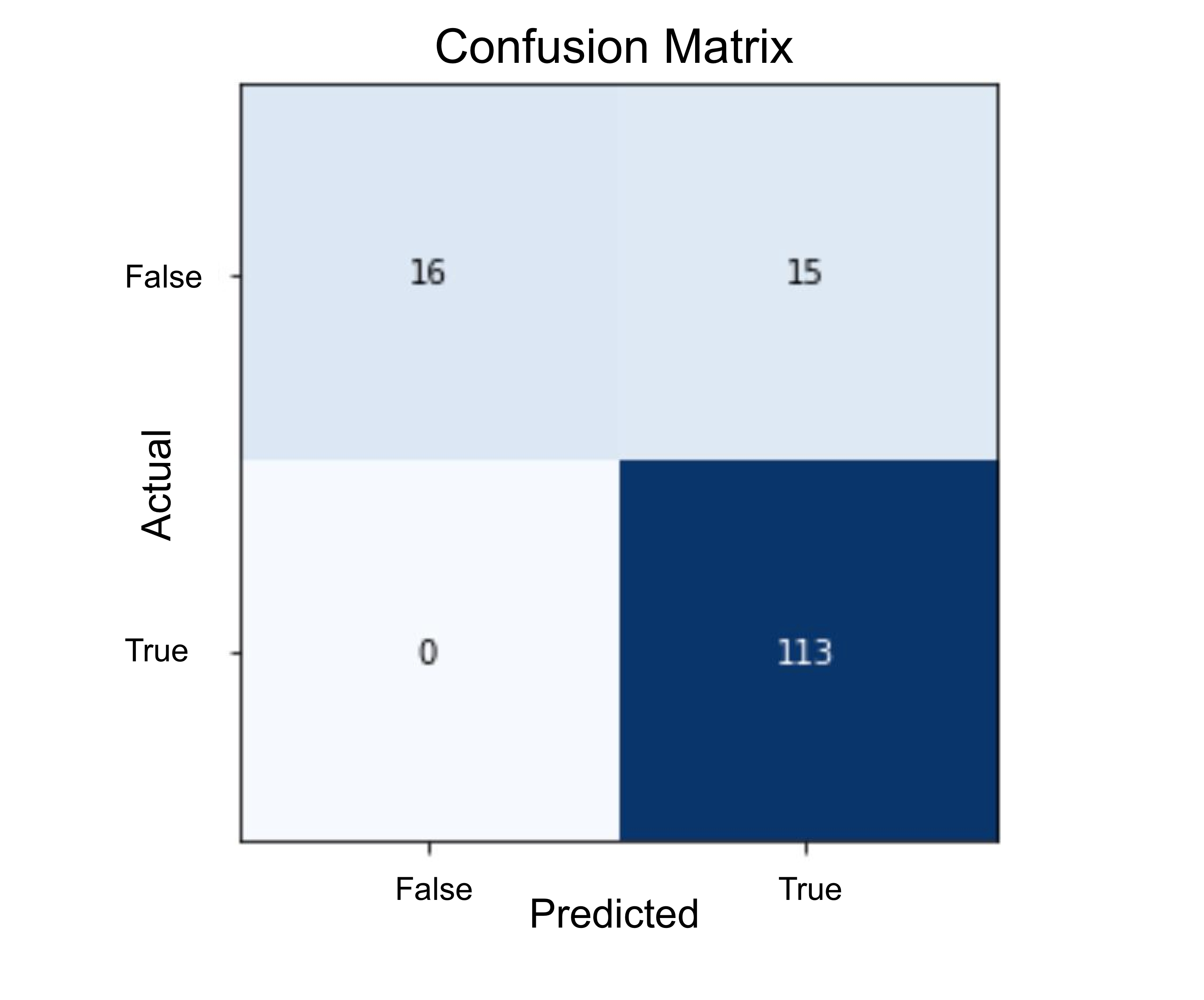}
  \caption{LSTM Confusion matrix}
  \label{fig:lstm_conf_matrix}
\end{figure}

This approach produced the highest recall. This means that this model was the best at catching the most amount of invalid surveys identified by the autoencoder. 

\subsection{Findings}
The LSTM approach was able to produce strong results, and it can certainly be used in an ensemble of multiple models to prevent general overfitting. Given its efficient runtime and high accuracy, we can also recommend it as a model of choice to predict autoencoder based labels if restrictions are posed. However, we ultimately stand by that the most generalizable results are achieved using a combination of approaches.

\section{HIDDEN MARKOV MODEL (HMM)}

Our third proposed method to determine the users' authenticity in survey responses is by analyzing the sequence of user movement using a Hidden Markov Model (HMM). HMM is an approach to model sequential data, and implies that the Markov Model underlying the data is unknown. Probabilistic graphical models such as HMM have been successfully used to identify user web activity. For such models, the sequences of observation are crucial for training and inference processes. We made a series of assumptions and data transformations, and we will provide an overview of the steps to produce the model and summary results and findings.

We converted the window aspect ratios into device types and discovered that certain users elected to take the survey on a laptop or mobile device. We believe that the movement patterns observed by people on mobile devices differ from those on a laptop. We solely focused on users who completed the survey using a laptop for modeling purposes.

We focused on users’ coordinates across the survey duration and discovered that there’s a lot of noise in the movements. To run an effective model, we converted the coordinates into discrete observations representing cardinal directions. For instance, a movement to the right of the x-axis and up on the y-axis is labeled as North East. In total, nine labels were created: North East, North West, North, South East, South West, South, West, East, and No Movement. Using these directions as states S, we create a sequence of observations concerning mouse movement activity by observing a user as they complete the survey. The priority is to understand the overall direction of the user movement.

We recognize that users are navigating through survey pages, so we use the coordinates of the next button to estimate when each user moves to the next page. After analyzing each survey page, we realized that each user has a unique layout and the mouse path that users exhibit varies. Furthermore, considering that the number of mouse movement records varies per page, we decided to analyze the first 200 observations per user. We also removed the users that took the survey multiple times. After multiple attempts those users have become accustomed to the survey design and movement would be based on memory.

 	Only 66 users met the defined criteria for further analysis in this approach. We trained the HMM using the Baum-Welch algorithm to estimate the transition matrix, state distribution, and output distribution. We train the algorithm to recognize the patterns in each page and apply the forward algorithm to calculate the observation log probability of each observed user sequence per page. A low log probability is interpreted as having a less likely occurrence. See table ~\ref{tab:sub_users} for an example of the results.

We scale each observation and apply an isolation forest to identify those suspicious users. Out of the 66 users, 11\%, or 7 users were labeled as suspicious. User Id: 422, 727, 866, 1272, 1297, 1314, 1495.

\begin{figure}
\centering
\begin{minipage}{.5\textwidth}
  \centering
  \includegraphics[width=\linewidth]{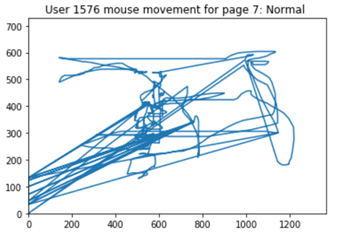}
  \caption{User 1576 mouse movement  for page 7: Normal}
  \label{fig:user1576}
\end{minipage}%
\begin{minipage}{.5\textwidth}
  \centering
  \includegraphics[width=\linewidth]{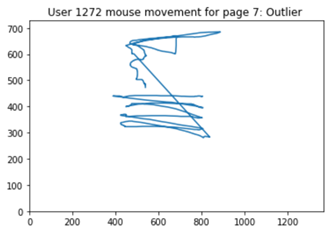}
  \caption{User 1272 mouse movement for page 7: Outlier}
  \label{fig:user1272}
\end{minipage}
\end{figure}

We compare two users for page 7 to illustrate their mouse movements. User 1576 movements move across the entire page (fig. ~\ref{fig:user1576}) while user 1272 movements are targeted and deliberate (fig. ~\ref{fig:user1272}).

\subsection{Assumptions and Limitations}

The accuracy of the HMM is dependent on the validity of the assumptions, and the quality of the data  \cite{Elbahi-16}, \cite{10.5555/3154538}. We therefore identify the assumptions and limitations of this approach.
\begin{itemize}
    \item The captured data doesn't distinguish when users are using their mouse to complete the survey vs browsing the internet.  
    \item The model assumes that the majority of users are completing the survey in good faith. If most users are falsely completing the survey, then the users that are attempting to complete the survey in good faith will be flagged.
    \item The model was trained on the first 200 sequential observations, and user’s patterns could differ as they progress through the pages. There are some users with 15,000 observations. Using an analogy, we are assuming that we can predict whether someone will win a 100m race using the first 10m. 
    \item The page labels were estimated using the coordinates of the next button on each page. Those labels represent our best estimate and may not truly reflect when the user page changes. 
\end{itemize}

\subsection{Findings} 
Despite the efficiency of such a probabilistic graphical model in segmenting and labeling stochastic sequences, its performance is adversely affected by the imperfect quality of data used for the construction of sequential observations. While the HMM can be useful in providing the probability of sequence, due to the quality of the data it shouldn't be the sole source. Therefore, we would suggest using a combination of methods in order to identify invalid survey responses.

\section{Conclusion}
To conclude, we have developed three different methods to validate psychometric survey responses for dotin Inc. These three methods helped us answer our initial research questions, in particular:
\begin{enumerate}
    \item{

Does the level of suspicious behavior vary across different types of survey questions? \\

From our outliers section, we were able to create general business rules to help us identify user behavior across pages.
\begin{itemize}
    \item Users that use the same scores across a single page can be flagged as suspicious.
    \item Users that take more than 5:30 minutes to answer the survey can be flagged as suspicious.
    \item Users that score above a standard deviation of 2 in our topic modeling, will be flagged as suspicious.
\end{itemize}

It is important to highlight the importance of having such business rules in the identification of suspicious behavior as flagging users could be an easy to implement rule approach to validating surveys. We envision this method to become the first line of defense from suspicious users, and an easy to implement solution to flag suspicious behavior across each page, and ultimately, the entire survey.
}

\item{
How do we use user mouse activity to validate survey answers to psychometric questions? \\

Through this analysis, we are looking to gain a better understanding of the user journey throughout the survey. The goal is to see if different ways of interacting with the survey could be a baseline to create a model that through direction and magnitude of mouse movement would help us identify whether a user is correctly filling out the survey.

To tackle the question we used both supervised and unsupervised techniques:
\begin{itemize}
    \item \textbf{Unsupervised/Supervised: LSTM} 
We implemented an autoencoder to generate an independent label, independent of dotin's current approach. We then used such variables as labels in an LSTM model that can classify suspicious user behavior.

    \item \textbf{Unsupervised: HMM}
We used a probabilistic approach that analyzed the sequence of user movement with the Hidden Markov Model and complemented it with the Isolation Forest Algorithm to find the number of suspicious users.

\end{itemize}
}
\end{enumerate}

Putting together our findings, we can now compare the performance and results generated by the three different methods:

\begin{table}
  \centering
  \caption{Methods comparison}
  \label{tab:commands}
  \begin{tabular}{|l|c|c|c|}
    \hline
                  & Rule-based Approach & LSTM & HMM \\
    \hline
    Method        & Rule-based & Supervised & Unsupervised \\
    Training time & NA & $\sim 6$ Minutes & $\sim 1$ Hour \\
    Percentage of Users Tested & 94\% & 44\% & 9\% \\
    Percentage of Suspicious Users & 44\% & 22\% & 11\% \\
    \hline
  \end{tabular}
\end{table}

As we can see, each model was trained on a different set of users due to the limitations we faced with the quality of the original data. Therefore we would not recommend using one single model at this point, yet we could proceed with a hybrid approach that takes into consideration 3 models to validate users.
We believe that an improved data collection method will further help improve the results of the individual models, as well as the overall hybrid model, enabling dotin Inc. improve the accuracy of their validation method for psychometric survey responses.

\section{Further work}
Extend data sets by results of new surveys, combine all 3 models together. Create a standard deviation score for each of the 3 approach, and use a threshold, lets say 5 to classify users.

\bibliographystyle{splncs04}
\bibliography{ms}

\newpage

\section{Appendix}
\appendix

\begin{table}
\section{Results of HMM for suspicious users}
  \caption{Suspicious users IDs with values by pages (P.)}
  \label{tab:sub_users}
  \resizebox{\textwidth}{!}{%
  \begin{tabular}{|c|c|c|c|c|c|c|c|c|c|c|c|c|c|c|c|}
    \hline
    ID & P. 1 & P. 2 & P. 3 & P. 4 & P. 5 & P. 6 & P. 7 & P. 8 & P. 9 & P. 10 & P. 11 & P. 12 & P. 13 & P. 14 & P. 15 \\
    \hline
    384 & 0.548 & 0.533 & -0.022 & -0.714 & 0.648 & 0.818 & 0.795 & 0.326 & 0.618 & 0.675 & 0.617 & 0.784 & 0.900 & 0.776 & 0.700 \\
    422 & -2.184 & -5.829 & -4.181 & -4.445 & -4.693 & -4.275 & -6.126 & -4.628 & -4.139 & -1.509 & -2.916 & -1.492 & -0.605 & -0.314 & -0.380 \\
    448 & 0.563 & 0.949 & 0.500 & -0.185 & 1.068 & 1.426 & 0.280 & 0.532 & 0.619 & 0.539 & -0.148 & 0.667 & -0.477 & 0.491 & 0.129 \\
    507 & 0.567 & 0.458 & -0.333 & 0.172 & -0.050 & 0.022 & 0.274 & -0.681 & -0.240 & 0.947 & 0.867 & 0.978 & 1.140 & 1.175 & 0.667 \\
    549 & 0.202 & 0.890 & 0.746 & 0.769 & 0.563 & 1.001 & 0.930 & 0.746 & 0.888 & 0.676 & 0.720 & -0.408 & -1.199 & -1.342 & -2.574 \\
    \hline
  \end{tabular}}
\end{table}

\begin{table}
  \section{All Created Features}
  \caption{Table of All Created Features}
  \label{tab:commands}
  \begin{tabular}{|p{1.5cm}|p{4cm}|p{8cm}|}
    \hline
    Feature Category &Feature Name & Feature Description\\
    \hline
Mouse Activity & Click Count & Number of times a user has answered a question \\
Mouse Activity & User Record Count & Number of times user has performed any mouse activity (scroll + moves + clicks) \\
Mouse Activity & Validation & Target Variable for supervised machine learning (boolean); classification modeling \\
Mouse Activity & average\_click\_delay & Average time taken between one click and the next; aggregated by user \\
Time & Max time lapsed & Total time taken by the user to complete the survey \\
Time & Time since last movement & Total time since the last mouse movement \\
Time & Time since last click & Total time since the last mouse click on a radio button \\
Time & Factor of difference & Quantify how the time it takes each user to complete the survey compares to expected read time calculations. \\
Distance & Total Distance & Total distance traveled by the user (Euclidean distance) \\
Distance & Measure\_width\_covered & A feature to give us a measure of screen coverage by user in terms of width (x coordinate) \\
Distance & Measure\_height\_covered & A feature to give us a measure of screen coverage by user in terms of height (y coordinate) \\
Direction & Moves left , Perc of left movements & The count and percentage of instances when the user moves from right to left on the screen \\
Direction & Moves right, perc of right movements & The count and percentage of instances when the user moves from left to right on the screen \\
Direction & Moves up, perc of up movements & The count and percentage of instances when the user moves from bottom to top on the screen \\
Direction & Moves down, perc of down movements & The count and percentage of instances when the user moves from top to bottom on the screen \\
Direction & No horizontal movement & Count and percentage of instances when user shows no horizontal movement on the screen \\
Direction & No vertical movement & Count and percentage of instances when user shows no vertical movement on the screen \\
Survey & Bf\_votes\_1,2,3,4,5, Bs\_votes\_1,2,3,4,5, Miq\_votes\_1,2,3,4,5, pgi\_votes\_1,2,3,4,5,6,7 & Choice of answers for each category for each question \\
Survey & Bf\_abs\_min\_max\_response, Bs\_abs\_min\_max\_response, Miq\_abs\_min\_max\_response, pgi\_abs\_min\_max\_response & Checks whether the user has selected all 1s (absolute minimum value of question choice selection) or 5s/7s (absolute max value of question choice selection) per question category type (bf\_questions, bs\_questions, miq\_questions, pgi\_questions). Boolean \\
Survey & Standard deviation on similar questions & Checks how user responses deviate on questions that  are similar in nature \\
    \hline
  \end{tabular}
\end{table}

\end{document}